\documentclass[twocolumn,pra,showpacs,superscriptaddress,english]{revtex4}

 \usepackage{amssymb}
 \usepackage{bm}
 \usepackage{array}
 \usepackage{graphicx}
 \usepackage{caption}
 \usepackage{amsmath}
 \usepackage{subfigure}
\usepackage[sort&compress]{natbib}

\begin{document}

\title{Phase noise and laser-cooling limits of optomechanical oscillators}
\author{Zhang-qi Yin}\email{yinzhangqi@gmail.com}
\affiliation{Department of Applied Physics, Xi'an Jiaotong
University, Xi'an 710049, China, and FOCUS center and MCTP,
Department of Physics, University of Michigan, Ann Arbor, Michigan
48109, USA}
\begin{abstract}
The noise from laser phase fluctuation sets a major technical
obstacle to cool the nano-mechanical oscillators to the quantum
region. We propose a cooling configuration based on the
opto-mechanical coupling with two cavity modes to significantly
reduce this phase noise by $(2\omega_m/\gamma)^2$ times, where
$\omega_m$ is the frequency of the mechanical mode and $\gamma$ is
the decay rate of the cavity mode. 
We also discuss the detection of the phonon number when the
mechanical oscillator is cooled near the quantum region and specify
the required conditions for this detection.
\end{abstract}

\pacs{42.50.Wk, 07.10.Cm, 03.65.Ta}
\maketitle

Cooling of the motion of nano-mechanical oscillators has attracted strong
interest recently \cite%
{2004Sci...304...74L,2004Natur.432.1002M,2006Natur.443..193N,2006Natur.444...71A}%
. When cooled to the quantum region, this system has many potential
applications, such as for mechanical
sensors\cite{2005PhT....58g..36S}, precision measurements
\cite{citeulike:3846325}, or quantum information processing
\cite{PhysRevLett.88.120401,vitali:030405}. The nano-mechanical
oscillators can be coupled to the cavity modes in optical resonators
and cooled through the sideband laser cooling
\cite{2007PhRvL..99i3901W,2007PhRvL..99i3902M}. For the sideband
cooling, the bandwidth of the cavity mode needs to be narrow
compared with the oscillation frequency of the mechanical oscillator
to resolve the sidebands
\cite{WilsonRae2008,2007PhRvL..99i3901W,2007PhRvL..99i3902M}.
Impressive experimental progress has been reported along this
direction,
which pushes the mean phonon number to the order of $100$ \cite%
{2008NatPhy..4..415S,citeulike:3884203,citeulike:3879646,PW09}. A
technical factor that limits the current temperature of the
oscillator is from the laser phase noise. The cooling laser is
typically red detuned from the cavity, and its inevitable phase
fluctuation will induce the photon number fluctuation in the cavity
mode. This fluctuation is equivalent to a thermal bath coupled to
the mechanical oscillator, and seriously limits the temperature of
the latter. If one assumes white noise model for the laser phase
fluctuation, to achieve the ground state cooling of the mechanical
oscillator, the result estimate has shown that the laser bandwidth
has to be extremely narrow, on the order of $10^{-4}-10^{-3}$ Hz,
which is almost impossible to achieve in this configuration
\cite{2008PhRvA..78b1801D}. When one takes into account the final
correlation time of the laser phase
fluctuation, this requirement gets significantly relaxed \cite%
{citeulike:4163560}. However, under practical laser bandwidth, the estimated
mean phonon number for the mechanical oscillator is still on the order of $%
10-100$ \cite{citeulike:4163560}, which is in agreement with the
experimental observation
\cite{citeulike:3884203,citeulike:3879646,PW09}. This shows that the
laser phase noise is still a major factor that limits the current
temperature of the mechanical oscillator in experiments.

In this paper, we propose a cooling configuration to significantly
reduce the influence of the laser phase noise. We exploit a
configuration where the mechanical oscillator is coupled to two
cavity modes, with the frequency splitting of the latter equal to
the mechanical oscillator frequency. A laser is resonantly driving
on the cavity mode with lower frequency. Because of anti-Stokes
scattering, phonons in mechanical oscillator are transformed into
photons in the other cavity mode with higher frequency. The photons
leak out of the cavity and the mechanical oscillator is cooled down.
If cavity decay rate $\gamma$ is much less than the mechanical
oscillator frequency $\omega_m$, the same cooling rate can be
realized with much lower driving power than single cavity mode
schemes. With a detailed calculation, we show that the phase noise
effects can be suppressed by $(2\omega_m /\gamma)^2$ times.
Besides, as long as the cooling laser driving strength $%
\Omega_c$ is less than mechanical frequency $\omega_m$, the laser
phase noise can be treated independent of the driving power. Similar
configurations have been investigated in order to generate
Einstein-Podolsky-Rosen (EPR) beams with very high entanglement in
the room temperature \cite{YH09}, to optimize the energy
transferring from phonon to photon in sideband cooling, to generate
entanglement between phonons and photons
\cite{2007arXiv0710.2383Z,2008PhRvA..78f3809M, 2008arXiv0812.3819M},
and to enhance the displacement sensitivity and the quantum
back-action of mechanical oscillator \cite{citeulike:4139733}.
Considering both phase noise and mechanical quality factor $Q$
induced cooling limits, we find that it is possible to cool the
mechanical oscillator down to the quantum regime by double cavity
modes scheme under the present experimental conditions. At last, we
discuss how to measure the mean thermal photon number of the
oscillator by measuring the blue and the red sideband spectra.
Similar to the sideband cooling of trapped ions
\cite{PhysRevLett.62.403,PhysRevA.62.053807}, there will be a large
imbalance between the blue and the red sideband output spectra, when
the mechanical mode is cooled down to the quantum regime
($\bar{n}_m< 1$).

\begin{figure}[tbph]
\centering
\includegraphics[width=7cm]{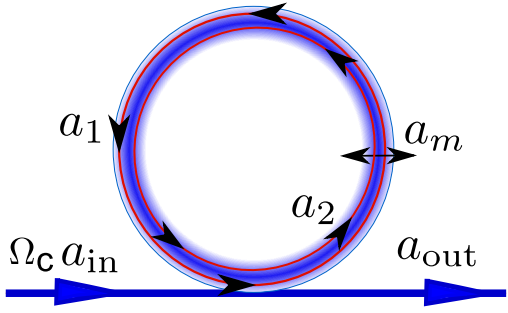}
\caption{(Color online) Double cavity mode scheme setup. There are
two cavity modes $a_{1}$ and $a_{2}$ couple with a mechanical mode
$a_{m}$.} \label{fig:scheme}
\end{figure}
As shown in Fig. \ref{fig:scheme}, there are two cavity modes $a_{1}$ and $%
a_{2}$ involving in the cooling process. The frequencies of the modes are $%
\omega _{1}$ and $\omega _{2}$, respectively. They are coupling with
a mechanical mode $a_{m}$ with frequency $\omega _{m}$. The
condition $\omega _{2}-\omega _{1}=\omega _{m}$ is fulfilled by
tuning either the mechanical mode frequency or the cavity mode
splitting. A laser is resonantly driving on the cavity mode $a_{1}$.
The present setup can be realized in Fabry Perot cavities , due to
the degeneracy of higher order modes \cite{2008PhRvA..78f3809M,
2008arXiv0812.3819M}, or in microsphere cavities whose closely
spaced azimuthal cavity modes or forward and backward cavity mode
splitting is tuned to match the mechanical
frequency \cite{TC09,Srin07,2008Dayan,YH09}. The Hamiltonian of the system is $%
H=H_{0}+H_{L}+H_{I}$ \cite%
{2007PhRvL..99i3901W,2007PhRvL..99i3902M,citeulike:4139733}, where
\begin{eqnarray}
H_{0} &=&-\Delta _{L}a_{1}^{\dagger }a_{1}+(-\Delta _{L}+\omega
_{m})a_{2}^{\dagger }a_{2}+\omega _{m}a_{m}^{\dagger }a_{m},  \label{eq:H1}
\\
H_{L} &=&\frac{\Omega _{c}e^{i\phi }}{2}(a_{1}+a_{2})+\mathrm{h.c.}, \\
H_{I} &=&\sum_{i,j=1,2}\eta \omega _{m}a_{i}^{\dagger }a_{j}(a_{m}^{\dagger
}+a_{m}).
\end{eqnarray}%
Here $\Omega _{c}$ is the driving strength of the cooling laser,
$\omega _{L}$ is the laser frequency, $\eta $ is the coupling parameter between
the cavity modes $%
a_{1,2}$ and the mechanical mode $a_{m}$, $\phi $ is the random phase noise \cite%
{2008PhRvA..78b1801D}. The dimensionless parameter $\eta $ is defined as $%
\eta =(\omega _{1}/\omega _{m})(x_{m}/R)$ , with $x_{m}=\sqrt{\hbar
/m\omega _{m}}$ as the zero-point motion of the mechanical resonator
mode $\omega _{m}$, $m$ as its effective mass, and $R$ as the cavity
radius. In typical system, the coupling constant $\eta $ is on the
order of $10^{-4}$. We denote detuning as $\Delta _{L}=\omega
_{L}-\omega _{1}$. The cavity modes and the mechanical mode are all
weakly dissipating with rates $\gamma _{1}$, $\gamma _{2}$ and
$\gamma _{m}$, which are much less than $\omega _{m}$. We get
quantum Langevin equations
\begin{equation}
\dot{a}_{j}=-i[a_{j},H]-\frac{\gamma _{j}}{2}a_{j}+\sqrt{\gamma _{j}}a_{j},~~%
\mathrm{for}~~j=1,2,m.  \label{eq:langvin}
\end{equation}

The driving and the decay terms in Eq. \eqref{eq:langvin} will be
balanced when time approaches infinity. The system approaches to a
classical steady state plus a quantum fluctuation. The latter one is
our main interest. To discuss the driving phase noise effects and
the quantum fluctuations, we apply transformations $a_{j}\rightarrow
a_{j}e^{-i\phi }$ and $a_{j}=\alpha _{j}+a_{j}$ for $j=1,2$,
$a_{m}=a_{m}+\beta $, respectively, where $\alpha _{j}$ and $\alpha
_{m}$ are the solutions of classical steady states, and $a_{j}$ and
$a_{m}$ are the quantum fluctuation operators. For the steady
states, the following conditions need to be fulfilled,
\begin{eqnarray*}
i\Delta _{L}\alpha _{1}-i\eta \omega _{m}(\alpha _{1}+\alpha _{2})(\beta
+\beta ^{\ast })-\frac{\gamma _{1}}{2}\alpha _{1}-i\frac{\Omega _{c}}{2}
&=&0, \\
-i(\omega _{m}-\Delta _{L})\alpha _{2}-i\eta \omega _{m}(\alpha
_{1}+\alpha _{2})(\beta +\beta ^{\ast }) \\-\frac{\gamma
_{2}}{2}\alpha _{2}-i\frac{\Omega _{c}}{2} &=&0,
\end{eqnarray*}%
where $\beta =-\eta |\alpha _{1}+\alpha _{2}|^{2}$, and $\Delta
_{L}=\eta \omega _{m}(\beta +\beta ^{\ast })$. Because $\omega
_{m}\gg \gamma _{1}$, it is easy to find that $|\alpha _{1}|\gg
|\alpha _{2}|$. We find that $\beta
\simeq -\eta |\alpha _{1}|^{2}$, $\alpha _{1}\simeq i\Omega _{c}/\gamma _{1}$%
, and $\alpha _{2}=(\Omega _{c}+2\eta ^{2}\omega _{m}\alpha _{1}|\alpha
_{1}|^{2})/(2i\omega _{m}+\gamma _{2})$. 
We find $\alpha _{1}/\alpha _{2}\simeq \gamma _{1}/(2\omega _{m})$. The
Langevin equations \eqref{eq:langvin} become 
\begin{equation}
\begin{aligned} \dot{a}_2=&-(i\omega_m+\frac{\gamma_2}{2}) a_2 -i\eta
\omega_m \alpha_1 (a^\dagger_m +a_m) +i\alpha_2 \dot{\phi}+
\sqrt{\gamma_2} a^{\mathrm{in}}_2,\\ \dot{a}_m=& -(i\omega_m
+\frac{\gamma_m}{2}) a_m -i\eta \omega_m (\alpha_1 a^\dagger_2 +
\alpha_1^* a_2) +\sqrt{\gamma_m} a_m^{\mathrm{in}}. \end{aligned}
\label{eq:langvin2}
\end{equation}%
In order to get Eq. \eqref{eq:langvin2}, at first we neglect $\alpha
_{2}$ terms in the coupling strength because it is much less than
$\alpha _{1}$. Then, as $\eta (\beta +\beta ^{\ast })\omega =2\eta
^{2}|\alpha _{1}|^{2}\omega _{m}\ll \omega _{m}$, we neglect the
coupling between $a_{1}$ and $a_{2}$.
As $\omega _{m}\gg \eta \omega _{m}$, there is no effective coupling between $%
a_{1}$ and $a_{m}$ modes. Therefore we neglect the $a_{1}$ mode in Eq. %
\eqref{eq:langvin}.

The phase noise term in Eqs. \eqref{eq:langvin2} induces the photon
number fluctuation, which heats the mechanical oscillator. Let us
briefly discuss the heating effects. In order to make the phase
noise effects more evident, we neglect the coupling between the
thermal bath and the mechanical oscillator. In the limit $\omega_m
\gg \gamma_2\gg \eta \omega_m \alpha_1$, we can adiabatically
eliminate the $a_2$ mode and get,
\begin{equation}  \label{eq:langvin3}
\dot{a}_m= - \frac{\tilde{\gamma}}{2} a_m + \sqrt{\tilde{\gamma}} a_2^{%
\mathrm{in}} - \sqrt{\tilde{\gamma}} \frac{\alpha_2}{\sqrt{\gamma_2}} \dot{%
\phi},
\end{equation}
where $\tilde{\gamma}=4\eta^2\omega_m^2|\alpha_1|^2/\gamma_2$. The
quantum
noise term $a^{\mathrm{in}}_2$ comes from the vacuum bath with correlation $%
\langle a^{\mathrm{in} \dagger}_2 (t) a^{\mathrm{in}}_2 (s) \rangle=\delta
(t-s)$. If we choose white noise model, the phase noise correlation is $%
\langle \dot{\phi}(t) \dot{\phi}(s) \rangle = 2\Gamma_L\delta(t-s)$, where $%
\Gamma_l$ is the linewidth of the driving laser
\cite{2008PhRvA..78b1801D}. We can treat the phase noise term
$\dot{\phi}$ the same as the vacuum noise term $a_2^\mathrm{in}$. In
order to cool the oscillator down to the ground state, we need make
sure that the heating strength of the phase noise term is much
less than the cooling effect of the vacuum noise term. Therefore we find $%
|\alpha_2|^2 \Gamma_l \ll \gamma_2$, where $|\alpha_2|^2 =n_2$ is
the mean photon number in the cavity mode $a_2$. This condition is
equivalent to the one in the single cavity mode cooling scheme
\cite{2008PhRvA..78b1801D}, with the mean cavity photon number
reduced by a factor $(\gamma_1/2\omega_m)^2$, leaving other
parameters unchanged. In resolved sideband regime, $\gamma_1$ is
much less than $\omega_m$. Therefore the phase noise heating effect
is suppressed by $(2\omega_m/\gamma_1)^2$ times. If $\Omega_c<
\omega_m$, the mean photon number in the cavity mode $a_2$ is less
than $1$. The ground state cooling condition becomes $\Gamma_l \ll
\gamma_2$, which is the same as the one used in the sideband cooling
of atoms. Besides, the same cooling rate can be realized by
$(\gamma_1/2\omega_m)^2$ times less the driving power than the
single cavity mode scheme, which is consistent with the results in
Ref. \cite{2008PhRvA..78f3809M} and \cite{2008arXiv0812.3819M}.

Now we briefly discuss the cooling limit related to the driving
phase noise and the mechanical quality $Q$ based on the current
experimental conditions.
The experimental available parameters are $\Gamma_l \simeq 10^3$ Hz, $%
\gamma_{1,2}/2\pi \sim 1$ MHz, and $\omega_m/2\pi \sim 100$ MHz \cite%
{2008NatPhy..4..415S}. Practically, $\alpha_2$ is much less than $%
\gamma_1/(\omega_m\eta) \sim 100$. We choose proper laser driving
power, which makes $|\alpha_2|^2< 10^3$. We find that
$|\alpha_2|^2\Gamma_l < \gamma_2$. So for the white noise model, the
limitation of thermal phonon number is below $1$, which is already
in the quantum regime. To be more
rigorous, we can choose Gaussian noise model with finite correlation time $%
\gamma^{-1}_c$ other than white noise model with zero correlation time \cite%
{citeulike:4163560}. The correlation function of the phase noise is $\langle
\dot{\phi}(t) \dot{\phi}(s) \rangle= \Gamma_l \gamma_c e^{-\gamma_c|t-s|}$.
In Ref. \cite{citeulike:4163560}, it was found that for the finite
correlation noise model, the effects of the phase noise reduces by $%
(\omega_m^2+\gamma_c^2)/\gamma_c^2$ times, compared with white noise
model. In the limit $\gamma_c \ll \omega_m$, we can conclude that
the phase noise effect is negligible at this time as
$|\alpha_2|^2\Gamma_l \gamma_c^2/(\omega_m^2 + \gamma_c^2) \ll
\gamma_2$.

The cooling limit is also related to the mechanical quality factor
$Q$. 
It is found that the limit of cooling is $n_{mf}>\gamma_m n_{mi} /
\gamma_2> n_{mi} /Q= k_BT/(\hbar\omega_m Q)$
\cite{WilsonRae2008,2007PhRvL..99i3902M}, where $T$ is the
environment temperature, $n_{mf}$ is the phonon number after laser
cooling, and $n_{mi}$ is the bath phonon number. In order to cool
oscillator to quantum regime, we should make sure that the initial
thermal phonon number ${n}_{mi}$ is much less than $Q$. Therefore,
it is necessary to either use high frequency and high quality $Q$
oscillators or cool the environment temperature before laser
cooling. Currently, the initial environment temperature is cooled
down to $1.65$ K and $Q$ is about $2000$ for mechanical oscillator
with frequency $\omega_m =62$ MHz \cite{citeulike:3879646}. So the
limit of $n_{mf}$ is $k_BT/(\hbar\omega_m Q)= 0.28$. It is also
found that $Q\sim \omega_m/T^3$ for very low temperature \cite%
{2009arXiv0901.1292A}. Therefore $Q$ is about $2\times 10^4$ for the
temperature around $600$ mK, which is still possible for $^3$He
cooling. The limit of $n_{mf}$ could be $0.01$. Combining the
cooling limit set up by phase noise effects and the mechanical
quality factor $Q$, we conclude that the present scheme greatly
decreases phase noise effects and makes cooling opto-mechanical
oscillator down to the quantum regime possible based on the current
experimental conditions.

\begin{figure}[htbp]
\centering
\includegraphics[width=7cm]{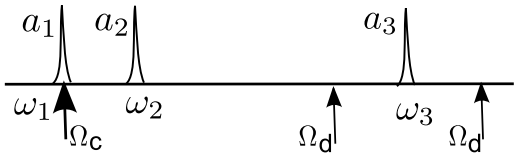}
\caption{Measurement setup.} \label{fig:measure}
\end{figure}

To verify the ground state cooling of the mechanical oscillator, we
need to directly measure the mean thermal phonon number ${n}_{mf}$.
Although the phonon number can be measured by displacement noise
spectrum
\cite{2008NatPhy..4..415S,citeulike:3884203,citeulike:3879646,PW09},
here we propose an other measurement scheme by measuring the output
light intensity. We will compare the two schemes later. As shown in
Fig. \ref{fig:measure}, we choose the third cavity mode $a_3$ with
frequency $\omega_3$. By weakly driving the red and the blue
detuning sidebands of the cavity mode $a_3$, we can measure the mean
thermal phonon number after the sideband cooling. The measurement
can be processed simultaneously and independently with the sideband
cooling. The
measurement scheme is similar to the one used in ion trap \cite%
{PhysRevLett.62.403,PhysRevA.62.053807}. However, in the present
setup, we need to make sure that the measurement process has
negligible effect on the cooling process. We will derive the
conditions of the driving laser strength. The Hamiltonian involved
with the measurement is
\begin{equation}  \label{eq:measure}
H_M=-\Delta_{L^{\prime }}n_3 + \omega_m n_m + (\frac{\Omega_d}{2} a_3 +
\mathrm{h.c.})+ \eta\omega_m n_3(a_m + a_m^\dagger),
\end{equation}
where $n_3=a_3^\dagger a_3$ and $n_m =a^\dagger_m a_m$, $\Omega_d$
is the driving strength of the detection laser, and
$\Delta_{L^{\prime }}= \omega_{L^{\prime }} -\omega_3$ is the
detuning between the driving laser and the
cavity mode $a_3$. We suppose that $a_3$ weakly decays with the rate $%
\gamma'_3$. The Langevin equations are similar to the Eq.
\eqref{eq:langvin}
by replacing $H$ with $H_M$. 
We apply the transformation $a_3 =a_3 + \alpha_3$ and $a_m = a_m
+\beta'$. The classical steady state satisfies
\begin{eqnarray*}
&&-\Delta_{L^{\prime }}\alpha_3 - i\eta\omega_m \alpha_3
(\beta'+{\beta'}^{*}) -
\frac{\gamma'_3}{2} \alpha_3 -i\frac{\Omega_d}{2} =0  \notag \\
&& \Delta_{L^{\prime }}+ 2\eta^2\omega_m |\alpha_3|^2 = -\omega_m,  \notag \\
&&\beta'=-\eta|\alpha_3|^2.
\end{eqnarray*}
Here we choose $\Delta_{L^{\prime }}+ 2\eta^2\omega_m |\alpha_3|^2=
-\omega_m $, which represents the blue sideband driving. In the limit $%
|\alpha_3|^2\gg |\langle a_3\rangle|^2$, we can linearize the
Langevin equations as
\begin{equation}  \label{eq:langvinMeasure}
\dot{a}_j= -i\eta\omega_m \alpha_3 (a_k+ a_k^\dagger) -i\omega_m a_j -\frac{%
\gamma'_j}{2} a_j +\sqrt{\gamma'_j}a_j^{\mathrm{in}},
\end{equation}
with $j,k=3,m$. Here we suppose that the mechanical oscillator
couples with an effective thermal bath with mean thermal number
$n_{mf}$ and effective coupling strength $\gamma'_{m}$ when laser
cooling is spontaneously processing. When the quantum regime
approaches, the effective coupling strength $\gamma'_{m}=\gamma_m+
\tilde{\gamma}$, where $\tilde{\gamma}$ is defined in Eq.
\eqref{eq:langvin3}. Before continuing, we need to make sure that
the classical steady state exists. Therefore the Routh-Hurwitz
criterion must be fulfilled \cite{2008PhRvA..78c2316G},
\begin{eqnarray*}
2\gamma'_m \gamma'_3 \{ (\gamma_3^{'2}+4\omega_m^{2})\gamma_3^{'2}+
\gamma'_m
[(\gamma'_m+ 2\gamma'_3)(\gamma_3^{'2}+ \omega_m^{'2}) \\
+ 2\gamma'_3 \omega_m^{2}]\} > \omega_m^2
(\eta|\alpha_3|\omega_m)^2(\gamma'_m+ 2\gamma'_3)^2.
\end{eqnarray*}
In the limit $\gamma'_m \ll \gamma'_3\ll \omega_m$, we find the condition is $%
2\gamma'_m\gamma'_3>\eta^2\omega_m^2|\alpha_3|^2$.

We change the energy reference by transformation $a_3\rightarrow
e^{-i\omega_mt} a_3$ and $a_m \rightarrow e^{-i\omega_mt}a_m$. In the limit $%
\omega_m \gg \gamma'_3,\gamma'_m, \eta\alpha_3\omega_m$, Eq. %
\eqref{eq:langvinMeasure} can be simplified by the rotating wave
approximation
\begin{equation}  \label{eq:langvinMeasure1}
\dot{a}_j = -i\eta\omega_m \alpha_3 a_k^\dagger -
\frac{\gamma'_j}{2} a_j + \sqrt{\gamma'_j} a_j^{\mathrm{in}},
\end{equation}
with $j,k=3,m$. We define the cavity or mechanical operator in the
frequency domain by Fourier transformation $a(t) =
\frac{1}{\sqrt{2\pi}} \int a (\omega)
e^{-i\omega(t-t_0)} d\omega$. 
With standard method \cite{QO}, we can solve the Langevin equations
\eqref{eq:langvinMeasure1} and get \begin{equation*}
a_3(\omega) = \frac{\sqrt{\gamma'_3}}{\Delta (\omega)} (\frac{\gamma'_m}{2}%
-i\omega) a_3^{\mathrm{in}}(\omega) + \frac{i\eta\omega_m\alpha_3}{
\Delta (\omega)}\sqrt{\gamma'_m} a_m^{\mathrm{in}\dagger} (-\omega).
\end{equation*}
where $\Delta(\omega) = (\frac{\gamma'_m}{2}-i\omega_m)(\frac{\gamma'_3}{2}%
-i\omega) -\eta^2\omega_m^2 \alpha_3^2$. We calculate the output
mode by the boundary condition $a_3^{\mathrm{in}} +
a_3^{\mathrm{out}} = \sqrt{\gamma'_3} a_3$,
\begin{equation*}
\begin{aligned}
a_3^{\mathrm{out}}(\omega) = \big[-1+ \frac{\gamma'_3}{\Delta} (\frac{\gamma'_m%
}{2} -i\omega)\big] a_3^{\mathrm{in}} (\omega)\\ +
\frac{i\eta\omega_m\alpha_3 \sqrt{\gamma'_m \gamma'_3}}{\Delta}
a_m^{\mathrm{in}\dagger} (-\omega).
\end{aligned}
\end{equation*}

We suppose that the mechanical oscillator is continuously cooled
when the measurement is processed. The cooling results can be
treated as an
effective thermal bath with mean phonon number $n_{mf}$. Therefore we have $%
\langle a_m^{ \mathrm{in}\dagger} (-\omega) a_m^{\mathrm{in}}
(\omega^{\prime }) \rangle=n_{mf} \delta(\omega-\omega)$ and
$\langle a_m^{ \mathrm{in}} (\omega) a_m^{\mathrm{in}\dagger}
(-\omega^{\prime })\rangle =(n_{mf}+1) \delta(\omega-\omega^{\prime
})$. The peak strength of the output field is
\begin{equation*}
I_b = \langle a^{\mathrm{out}\dagger}_3 (0)
a^{\mathrm{out}}_3(0)\rangle = \frac{\eta^2 \omega_m^2
\alpha_3^2}{\Delta^2(0)} \gamma'_m \gamma'_3 (n_{mf} + 1).
\end{equation*}
Similarly, if we choose diving at the red sideband with
$\Delta_{L^{\prime }} + 2\eta^2 \omega_m |\alpha_3|^2 = \omega_m$
and with the same driving power, the peak strength of the output
field is
\begin{equation*}
I_r = \langle a^{\mathrm{out}\dagger}_3 (0)
a^{\mathrm{out}}_3(0)\rangle = \frac{\eta^2 \omega_m^2
\alpha_3^2}{\Delta^{\prime 2}(0)} \gamma'_m \gamma'_3 n_{mf}.
\end{equation*}
where $\Delta^{\prime }(\omega) = (\frac{\gamma'_m}{2}-i\omega )(\frac{%
\gamma'_3}{2}-i\omega) +\eta^2\omega_m^2 \alpha_3^2$. In the limit $%
(\eta\omega_m\alpha_3)^2 \ll \gamma'_m\gamma'_3/8$ (the stable condition $%
2\gamma'_m\gamma'_3>\eta^2\omega_m^2|\alpha_3|^2$ is automatically
fulfilled), we get $\Delta(0)\simeq \Delta^{\prime }(0)$. The ratio
between the red and the blue sideband output central peak strengths
is $I_r/I_b = n_{mf} /(n_{mf} +1)$. Therefore, we can measure the
final thermal phonon number by measuring the ratio of two sideband
field strength. If we can cool the mechanical mode
to the ground state with $n_{mf} \rightarrow 0$, we will find that the ratio $%
I_r/I_b$ approaches zero. $\alpha_3$ is on the order of $%
10$ for practical parameters \cite{2008NatPhy..4..415S}, which is
much less than the cooling field amplitude $\alpha_1 \sim 10^3$ or
more. Therefore the measurement has negligible effects on the
cooling process.

Before conclusion, we compare the thermal phonon measurement schemes
between ours and those used in the current experiments
\cite{2008NatPhy..4..415S,citeulike:3884203,citeulike:3879646,PW09}.
The currently used measurement schemes compare the initial and the
final displacement noise spectra and get the final thermal phonon
number. Therefore the bath temperature is needed to calculate the
final thermal phonon number. The measurement precision is related to
the bath temperature measurement and noise spectrum measurement
precision \cite{PW09}. Because of background noise, the scheme is
less and less precise when the system approaches the quantum regime.
In the scheme that we proposed in this paper, both the noise
spectrum and the bath temperature are not needed to get $n_{mf}$. We
only need to measure the output red and blue sideband intensity
spectra. Besides, our scheme is reliable only when $n_{mf}$ is
comparable to $1$. Therefore, our scheme is much more accurate in
the quantum regime than in the classical regime. In fact, the large
imbalance between the red and the blue sideband spectra is the
direct signal that the oscillator is cooled down to the quantum
regime.

In conclusion, we have proposed a double cavity modes scheme to
eliminate the driving phase noise in sideband cooling of
opto-mechanical oscillators. We show that phase noise effects are
suppressed by $(2\omega_m/\gamma)^2$ times. The cooling limit from
the laser phase noise is already in the quantum regime for the
present experimental parameters. Combining the limits by the phase
noise and the mechanical quality factor $Q$, we conclude that it is
possible to cool down to the quantum regime at present. At last, we
discuss how to detect the thermal phonon number by measuring the red
and the blue sideband spectra when the mechanical oscillator is
cooled near the quantum regime. We specify the required conditions
for this measurement. We compare the measurement scheme to the
currently used ones.

We thank Lu-ming Duan for helpful discussions and valuable comments
on the paper, and Xiao-shun Jiang and Tongcang Li for useful
suggestions. ZY was supported by the Government of China through CSC
(Contract No. 2007102530).

{\em Note added}: After submitting the paper, We found Ref.
\cite{2007arXiv0710.2383Z} was published in \cite{ZJ+09}, where
rigorous discussion on suppressing the phase noise by double cavity
modes scheme was added.


\begin{thebibliography}{31}
\expandafter\ifx\csname
natexlab\endcsname\relax\def\natexlab#1{#1}\fi
\expandafter\ifx\csname bibnamefont\endcsname\relax
  \def\bibnamefont#1{#1}\fi
\expandafter\ifx\csname bibfnamefont\endcsname\relax
  \def\bibfnamefont#1{#1}\fi
\expandafter\ifx\csname citenamefont\endcsname\relax
  \def\citenamefont#1{#1}\fi
\expandafter\ifx\csname url\endcsname\relax
  \def\url#1{\texttt{#1}}\fi
\expandafter\ifx\csname urlprefix\endcsname\relax\def\urlprefix{URL
}\fi \providecommand{\bibinfo}[2]{#2}
\providecommand{\eprint}[2][]{\url{#2}}

\bibitem[{\citenamefont{{LaHaye} et~al.}(2004)\citenamefont{{LaHaye}, {Buu},
  {Camarota}, and {Schwab}}}]{2004Sci...304...74L}
\bibinfo{author}{\bibfnamefont{M.~D.} \bibnamefont{{LaHaye}}},
  \bibinfo{author}{\bibfnamefont{O.}~\bibnamefont{{Buu}}},
  \bibinfo{author}{\bibfnamefont{B.}~\bibnamefont{{Camarota}}},
  \bibnamefont{and} \bibinfo{author}{\bibfnamefont{K.~C.}
  \bibnamefont{{Schwab}}}, \bibinfo{journal}{Science}
  \textbf{\bibinfo{volume}{304}}, \bibinfo{pages}{74} (\bibinfo{year}{2004}).

\bibitem[{\citenamefont{{Metzger} and {Karrai}}(2004)}]{2004Natur.432.1002M}
\bibinfo{author}{\bibfnamefont{C.~H.} \bibnamefont{{Metzger}}}
  \bibnamefont{and} \bibinfo{author}{\bibfnamefont{K.}~\bibnamefont{{Karrai}}},
  \bibinfo{journal}{\nat} \textbf{\bibinfo{volume}{432}}, \bibinfo{pages}{1002}
  (\bibinfo{year}{2004}).

\bibitem[{\citenamefont{{Naik} et~al.}(2006)\citenamefont{{Naik}, {Buu},
  {Lahaye}, {Armour}, {Clerk}, {Blencowe}, and {Schwab}}}]{2006Natur.443..193N}
\bibinfo{author}{\bibfnamefont{A.}~\bibnamefont{{Naik}}},
  \bibinfo{author}{\bibfnamefont{O.}~\bibnamefont{{Buu}}},
  \bibinfo{author}{\bibfnamefont{M.~D.} \bibnamefont{{Lahaye}}},
  \bibinfo{author}{\bibfnamefont{A.~D.} \bibnamefont{{Armour}}},
  \bibinfo{author}{\bibfnamefont{A.~A.} \bibnamefont{{Clerk}}},
  \bibinfo{author}{\bibfnamefont{M.~P.} \bibnamefont{{Blencowe}}},
  \bibnamefont{and} \bibinfo{author}{\bibfnamefont{K.~C.}
  \bibnamefont{{Schwab}}}, \bibinfo{journal}{\nat}
  \textbf{\bibinfo{volume}{443}}, \bibinfo{pages}{193} (\bibinfo{year}{2006}).

\bibitem[{\citenamefont{{Arcizet} et~al.}(2006)\citenamefont{{Arcizet},
  {Cohadon}, {Briant}, {Pinard}, and {Heidmann}}}]{2006Natur.444...71A}
\bibinfo{author}{\bibfnamefont{O.}~\bibnamefont{{Arcizet}}},
  \bibinfo{author}{\bibfnamefont{P.-F.} \bibnamefont{{Cohadon}}},
  \bibinfo{author}{\bibfnamefont{T.}~\bibnamefont{{Briant}}},
  \bibinfo{author}{\bibfnamefont{M.}~\bibnamefont{{Pinard}}}, \bibnamefont{and}
  \bibinfo{author}{\bibfnamefont{A.}~\bibnamefont{{Heidmann}}},
  \bibinfo{journal}{\nat} \textbf{\bibinfo{volume}{444}}, \bibinfo{pages}{71}
  (\bibinfo{year}{2006}).

\bibitem[{\citenamefont{{Schwab} and {Roukes}}(2005)}]{2005PhT....58g..36S}
\bibinfo{author}{\bibfnamefont{K.~C.} \bibnamefont{{Schwab}}} \bibnamefont{and}
  \bibinfo{author}{\bibfnamefont{M.~L.} \bibnamefont{{Roukes}}},
  \bibinfo{journal}{Physics Today} \textbf{\bibinfo{volume}{58}},
  \bibinfo{pages}{36} (\bibinfo{year}{2005}).

\bibitem[{\citenamefont{Kippenberg and Vahala}(2008)}]{citeulike:3846325}
\bibinfo{author}{\bibfnamefont{T.~J.} \bibnamefont{Kippenberg}}
  \bibnamefont{and} \bibinfo{author}{\bibfnamefont{K.~J.}
  \bibnamefont{Vahala}}, \bibinfo{journal}{Science}
  \textbf{\bibinfo{volume}{321}}, \bibinfo{pages}{1172} (\bibinfo{year}{2008}).

\bibitem[{\citenamefont{Mancini et~al.}(2002)\citenamefont{Mancini,
  Giovannetti, Vitali, and Tombesi}}]{PhysRevLett.88.120401}
\bibinfo{author}{\bibfnamefont{S.}~\bibnamefont{Mancini}},
  \bibinfo{author}{\bibfnamefont{V.}~\bibnamefont{Giovannetti}},
  \bibinfo{author}{\bibfnamefont{D.}~\bibnamefont{Vitali}}, \bibnamefont{and}
  \bibinfo{author}{\bibfnamefont{P.}~\bibnamefont{Tombesi}},
  \bibinfo{journal}{Phys. Rev. Lett.} \textbf{\bibinfo{volume}{88}},
  \bibinfo{pages}{120401} (\bibinfo{year}{2002}).

\bibitem[{\citenamefont{Vitali et~al.}(2007)\citenamefont{Vitali, Gigan,
  Ferreira, B\"{o}hm, Tombesi, Guerreiro, Vedral, Zeilinger, and
  Aspelmeyer}}]{vitali:030405}
\bibinfo{author}{\bibfnamefont{D.}~\bibnamefont{Vitali}},
  \bibinfo{author}{\bibfnamefont{S.}~\bibnamefont{Gigan}},
  \bibinfo{author}{\bibfnamefont{A.}~\bibnamefont{Ferreira}},
  \bibinfo{author}{\bibfnamefont{H.~R.} \bibnamefont{B\"{o}hm}},
  \bibinfo{author}{\bibfnamefont{P.}~\bibnamefont{Tombesi}},
  \bibinfo{author}{\bibfnamefont{A.}~\bibnamefont{Guerreiro}},
  \bibinfo{author}{\bibfnamefont{V.}~\bibnamefont{Vedral}},
  \bibinfo{author}{\bibfnamefont{A.}~\bibnamefont{Zeilinger}},
  \bibnamefont{and}
  \bibinfo{author}{\bibfnamefont{M.}~\bibnamefont{Aspelmeyer}},
  \bibinfo{journal}{Phys. Rev. Lett.} \textbf{\bibinfo{volume}{98}},
  \bibinfo{pages}{030405} (\bibinfo{year}{2007}).

\bibitem[{\citenamefont{{Wilson-Rae} et~al.}(2007)\citenamefont{{Wilson-Rae},
  {Nooshi}, {Zwerger}, and {Kippenberg}}}]{2007PhRvL..99i3901W}
\bibinfo{author}{\bibfnamefont{I.}~\bibnamefont{{Wilson-Rae}}},
  \bibinfo{author}{\bibfnamefont{N.}~\bibnamefont{{Nooshi}}},
  \bibinfo{author}{\bibfnamefont{W.}~\bibnamefont{{Zwerger}}},
  \bibnamefont{and} \bibinfo{author}{\bibfnamefont{T.~J.}
  \bibnamefont{{Kippenberg}}}, \bibinfo{journal}{Phys. Rev. Lett.}
  \textbf{\bibinfo{volume}{99}}, \bibinfo{pages}{093901}
  (\bibinfo{year}{2007}).

\bibitem[{\citenamefont{{Marquardt} et~al.}(2007)\citenamefont{{Marquardt},
  {Chen}, {Clerk}, and {Girvin}}}]{2007PhRvL..99i3902M}
\bibinfo{author}{\bibfnamefont{F.}~\bibnamefont{{Marquardt}}},
  \bibinfo{author}{\bibfnamefont{J.~P.} \bibnamefont{{Chen}}},
  \bibinfo{author}{\bibfnamefont{A.~A.} \bibnamefont{{Clerk}}},
  \bibnamefont{and} \bibinfo{author}{\bibfnamefont{S.~M.}
  \bibnamefont{{Girvin}}}, \bibinfo{journal}{Phys. Rev. Lett.}
  \textbf{\bibinfo{volume}{99}}, \bibinfo{pages}{093902}
  (\bibinfo{year}{2007}).

\bibitem[{\citenamefont{Wilson-Rae et~al.}(2008)\citenamefont{Wilson-Rae,
  Nooshi, Dobrindt, Kippenberg, and Zwerger}}]{WilsonRae2008}
\bibinfo{author}{\bibfnamefont{I.}~\bibnamefont{Wilson-Rae}},
  \bibinfo{author}{\bibfnamefont{N.}~\bibnamefont{Nooshi}},
  \bibinfo{author}{\bibfnamefont{J.}~\bibnamefont{Dobrindt}},
  \bibinfo{author}{\bibfnamefont{T.~J.} \bibnamefont{Kippenberg}},
  \bibnamefont{and} \bibinfo{author}{\bibfnamefont{W.}~\bibnamefont{Zwerger}},
  \bibinfo{journal}{New J. Phys.} \textbf{\bibinfo{volume}{10}},
  \bibinfo{pages}{095007} (\bibinfo{year}{2008}).

\bibitem[{\citenamefont{{Schliesser} et~al.}(2008)\citenamefont{{Schliesser},
  {Rivi{\`e}re}, {Anetsberger}, {Arcizet}, and
  {Kippenberg}}}]{2008NatPhy..4..415S}
\bibinfo{author}{\bibfnamefont{A.}~\bibnamefont{{Schliesser}}},
  \bibinfo{author}{\bibfnamefont{R.}~\bibnamefont{{Rivi{\`e}re}}},
  \bibinfo{author}{\bibfnamefont{G.}~\bibnamefont{{Anetsberger}}},
  \bibinfo{author}{\bibfnamefont{O.}~\bibnamefont{{Arcizet}}},
  \bibnamefont{and} \bibinfo{author}{\bibfnamefont{T.~J.}
  \bibnamefont{{Kippenberg}}}, \bibinfo{journal}{Nature Physics}
  \textbf{\bibinfo{volume}{4}}, \bibinfo{pages}{415} (\bibinfo{year}{2008}).

\bibitem[{\citenamefont{Gr\"oblacher et~al.}(2009)\citenamefont{Gr\"oblacher,
  Hertzberg, Vanner, Gigan, Schwab, and Aspelmeyer}}]{citeulike:3884203}
\bibinfo{author}{\bibfnamefont{S.}~\bibnamefont{Gr\"oblacher}},
  \bibinfo{author}{\bibfnamefont{J.~B.} \bibnamefont{Hertzberg}},
  \bibinfo{author}{\bibfnamefont{M.~R.} \bibnamefont{Vanner}},
  \bibinfo{author}{\bibfnamefont{S.}~\bibnamefont{Gigan}},
  \bibinfo{author}{\bibfnamefont{K.~C.} \bibnamefont{Schwab}},
  \bibnamefont{and}
  \bibinfo{author}{\bibfnamefont{M.}~\bibnamefont{Aspelmeyer}},
  \bibinfo{journal}{Nature Physics} \textbf{\bibinfo{volume}{5}},
  \bibinfo{pages}{485} (\bibinfo{year}{2009}).

\bibitem[{\citenamefont{Schliesser et~al.}(2009)\citenamefont{Schliesser,
  Arcizet, Rivi\`ere, Anetsberger, and Kippenberg}}]{citeulike:3879646}
\bibinfo{author}{\bibfnamefont{A.}~\bibnamefont{Schliesser}},
  \bibinfo{author}{\bibfnamefont{O.}~\bibnamefont{Arcizet}},
  \bibinfo{author}{\bibfnamefont{R.}~\bibnamefont{Rivi\`ere}},
  \bibinfo{author}{\bibfnamefont{G.}~\bibnamefont{Anetsberger}},
  \bibnamefont{and} \bibinfo{author}{\bibfnamefont{T.~J.}
  \bibnamefont{Kippenberg}}, \bibinfo{journal}{Nature Physics}
  \textbf{\bibinfo{volume}{5}}, \bibinfo{pages}{509} (\bibinfo{year}{2009}).

\bibitem[{\citenamefont{Park and Wang}(2009)}]{PW09}
\bibinfo{author}{\bibfnamefont{Y.-S.} \bibnamefont{Park}} \bibnamefont{and}
  \bibinfo{author}{\bibfnamefont{H.}~\bibnamefont{Wang}},
  \bibinfo{journal}{Nature Physics} \textbf{\bibinfo{volume}{5}},
  \bibinfo{pages}{489} (\bibinfo{year}{2009}).

\bibitem[{\citenamefont{{Di{\'o}si}}(2008)}]{2008PhRvA..78b1801D}
\bibinfo{author}{\bibfnamefont{L.}~\bibnamefont{{Di{\'o}si}}},
  \bibinfo{journal}{\pra} \textbf{\bibinfo{volume}{78}},
  \bibinfo{pages}{021801(R)} (\bibinfo{year}{2008}).

\bibitem[{\citenamefont{Rabl et~al.}(2009)\citenamefont{Rabl, Genes, Hammerer,
  and Aspelmeyer}}]{citeulike:4163560}
\bibinfo{author}{\bibfnamefont{P.}~\bibnamefont{Rabl}},
  \bibinfo{author}{\bibfnamefont{C.}~\bibnamefont{Genes}},
  \bibinfo{author}{\bibfnamefont{K.}~\bibnamefont{Hammerer}}, \bibnamefont{and}
  \bibinfo{author}{\bibfnamefont{M.}~\bibnamefont{Aspelmeyer}}
  (\bibinfo{year}{2009}), \eprint{arXiv:0903.1637}.

\bibitem[{\citenamefont{{Zhang-qi Yin} and {Y.-J. Han}}(2009)}]{YH09}
\bibinfo{author}{\bibnamefont{{Zhang-qi Yin}}} \bibnamefont{and}
  \bibinfo{author}{\bibnamefont{{Y.-J. Han}}}, \bibinfo{journal}{Phys. Rev. A}
  \textbf{\bibinfo{volume}{79}}, \bibinfo{pages}{024301}
  (\bibinfo{year}{2009}).

\bibitem[{\citenamefont{{Zhao} et~al.}(2009)\citenamefont{{Zhao}, {Ju}, {Miao},
  {Gras}, {Fan}, and {Blair}}}]{2007arXiv0710.2383Z}
\bibinfo{author}{\bibfnamefont{C.}~\bibnamefont{{Zhao}}},
  \bibinfo{author}{\bibfnamefont{L.}~\bibnamefont{{Ju}}},
  \bibinfo{author}{\bibfnamefont{H.}~\bibnamefont{{Miao}}},
  \bibinfo{author}{\bibfnamefont{S.}~\bibnamefont{{Gras}}},
  \bibinfo{author}{\bibfnamefont{Y.}~\bibnamefont{{Fan}}}, \bibnamefont{and}
  \bibinfo{author}{\bibfnamefont{D.~G.} \bibnamefont{{Blair}}}
  (\bibinfo{year}{2009}), \eprint{arXiv:0710.2383v3}.

\bibitem[{\citenamefont{{Miao} et~al.}(2008)\citenamefont{{Miao}, {Zhao}, {Ju},
  {Gras}, {Barriga}, {Zhang}, and {Blair}}}]{2008PhRvA..78f3809M}
\bibinfo{author}{\bibfnamefont{H.}~\bibnamefont{{Miao}}},
  \bibinfo{author}{\bibfnamefont{C.}~\bibnamefont{{Zhao}}},
  \bibinfo{author}{\bibfnamefont{L.}~\bibnamefont{{Ju}}},
  \bibinfo{author}{\bibfnamefont{S.}~\bibnamefont{{Gras}}},
  \bibinfo{author}{\bibfnamefont{P.}~\bibnamefont{{Barriga}}},
  \bibinfo{author}{\bibfnamefont{Z.}~\bibnamefont{{Zhang}}}, \bibnamefont{and}
  \bibinfo{author}{\bibfnamefont{D.~G.} \bibnamefont{{Blair}}},
  \bibinfo{journal}{\pra} \textbf{\bibinfo{volume}{78}},
  \bibinfo{pages}{063809} (\bibinfo{year}{2008}).

\bibitem[{\citenamefont{Miao et~al.}(2009)\citenamefont{Miao, Zhao, Ju, and
  Blair}}]{2008arXiv0812.3819M}
\bibinfo{author}{\bibfnamefont{H.}~\bibnamefont{Miao}},
  \bibinfo{author}{\bibfnamefont{C.}~\bibnamefont{Zhao}},
  \bibinfo{author}{\bibfnamefont{L.}~\bibnamefont{Ju}}, \bibnamefont{and}
  \bibinfo{author}{\bibfnamefont{D.~G.} \bibnamefont{Blair}},
  \bibinfo{journal}{Phys. Rev. A} \textbf{\bibinfo{volume}{79}},
  \bibinfo{eid}{063801} (\bibinfo{year}{2009}).

\bibitem[{\citenamefont{Dobrindt and Kippenberg}(2009)}]{citeulike:4139733}
\bibinfo{author}{\bibfnamefont{J.~M.} \bibnamefont{Dobrindt}} \bibnamefont{and}
  \bibinfo{author}{\bibfnamefont{T.~J.} \bibnamefont{Kippenberg}}
  (\bibinfo{year}{2009}), \eprint{arXiv:0903.1013}.

\bibitem[{\citenamefont{Diedrich et~al.}(1989)\citenamefont{Diedrich,
  Bergquist, Itano, and Wineland}}]{PhysRevLett.62.403}
\bibinfo{author}{\bibfnamefont{F.}~\bibnamefont{Diedrich}},
  \bibinfo{author}{\bibfnamefont{J.~C.} \bibnamefont{Bergquist}},
  \bibinfo{author}{\bibfnamefont{W.~M.} \bibnamefont{Itano}}, \bibnamefont{and}
  \bibinfo{author}{\bibfnamefont{D.~J.} \bibnamefont{Wineland}},
  \bibinfo{journal}{Phys. Rev. Lett.} \textbf{\bibinfo{volume}{62}},
  \bibinfo{pages}{403} (\bibinfo{year}{1989}).

\bibitem[{\citenamefont{Turchette et~al.}(2000)\citenamefont{Turchette, Myatt,
  King, Sackett, Kielpinski, Itano, Monroe, and Wineland}}]{PhysRevA.62.053807}
\bibinfo{author}{\bibfnamefont{Q.~A.} \bibnamefont{Turchette}},
  \bibinfo{author}{\bibfnamefont{C.~J.} \bibnamefont{Myatt}},
  \bibinfo{author}{\bibfnamefont{B.~E.} \bibnamefont{King}},
  \bibinfo{author}{\bibfnamefont{C.~A.} \bibnamefont{Sackett}},
  \bibinfo{author}{\bibfnamefont{D.}~\bibnamefont{Kielpinski}},
  \bibinfo{author}{\bibfnamefont{W.~M.} \bibnamefont{Itano}},
  \bibinfo{author}{\bibfnamefont{C.}~\bibnamefont{Monroe}}, \bibnamefont{and}
  \bibinfo{author}{\bibfnamefont{D.~J.} \bibnamefont{Wineland}},
  \bibinfo{journal}{Phys. Rev. A} \textbf{\bibinfo{volume}{62}},
  \bibinfo{pages}{053807} (\bibinfo{year}{2000}).

\bibitem[{\citenamefont{Tomes and Carmon}(2009)}]{TC09}
\bibinfo{author}{\bibfnamefont{M.}~\bibnamefont{Tomes}} \bibnamefont{and}
  \bibinfo{author}{\bibfnamefont{T.}~\bibnamefont{Carmon}},
  \bibinfo{journal}{Phys. Rev. Lett.} \textbf{\bibinfo{volume}{102}},
  \bibinfo{eid}{113601} (\bibinfo{year}{2009}).

\bibitem[{\citenamefont{Srinivasan and Painter}(2007)}]{Srin07}
\bibinfo{author}{\bibfnamefont{K.}~\bibnamefont{Srinivasan}} \bibnamefont{and}
  \bibinfo{author}{\bibfnamefont{O.}~\bibnamefont{Painter}},
  \bibinfo{journal}{Phys. Rev. A} \textbf{\bibinfo{volume}{75}},
  \bibinfo{pages}{023814} (\bibinfo{year}{2007}).

\bibitem[{\citenamefont{{Dayan} et~al.}(2008)\citenamefont{{Dayan}, {Parkins},
  {Aoki}, {Ostby}, {Vahala}, and {Kimble}}}]{2008Dayan}
\bibinfo{author}{\bibfnamefont{B.}~\bibnamefont{{Dayan}}},
  \bibinfo{author}{\bibfnamefont{A.~S.} \bibnamefont{{Parkins}}},
  \bibinfo{author}{\bibfnamefont{T.}~\bibnamefont{{Aoki}}},
  \bibinfo{author}{\bibfnamefont{E.~P.} \bibnamefont{{Ostby}}},
  \bibinfo{author}{\bibfnamefont{K.~J.} \bibnamefont{{Vahala}}},
  \bibnamefont{and} \bibinfo{author}{\bibfnamefont{H.~J.}
  \bibnamefont{{Kimble}}}, \bibinfo{journal}{Science}
  \textbf{\bibinfo{volume}{319}}, \bibinfo{pages}{1062} (\bibinfo{year}{2008}).

\bibitem[{\citenamefont{{Arcizet} et~al.}(2009)\citenamefont{{Arcizet},
  {Rivi{\`e}re}, {Schliesser}, {Anetsberger}, and
  {Kippenberg}}}]{2009arXiv0901.1292A}
\bibinfo{author}{\bibfnamefont{O.}~\bibnamefont{{Arcizet}}},
  \bibinfo{author}{\bibfnamefont{R.}~\bibnamefont{{Rivi{\`e}re}}},
  \bibinfo{author}{\bibfnamefont{A.}~\bibnamefont{{Schliesser}}},
  \bibinfo{author}{\bibfnamefont{G.}~\bibnamefont{{Anetsberger}}},
  \bibnamefont{and} \bibinfo{author}{\bibfnamefont{T.~J.}
  \bibnamefont{{Kippenberg}}}, \bibinfo{journal}{Phys. Rev. A}
  \textbf{\bibinfo{volume}{80}}, \bibinfo{pages}{021803(R)}
  (\bibinfo{year}{2009}).

\bibitem[{\citenamefont{{Genes} et~al.}(2008)\citenamefont{{Genes}, {Mari},
  {Tombesi}, and {Vitali}}}]{2008PhRvA..78c2316G}
\bibinfo{author}{\bibfnamefont{C.}~\bibnamefont{{Genes}}},
  \bibinfo{author}{\bibfnamefont{A.}~\bibnamefont{{Mari}}},
  \bibinfo{author}{\bibfnamefont{P.}~\bibnamefont{{Tombesi}}},
  \bibnamefont{and} \bibinfo{author}{\bibfnamefont{D.}~\bibnamefont{{Vitali}}},
  \bibinfo{journal}{\pra} \textbf{\bibinfo{volume}{78}},
  \bibinfo{pages}{032316} (\bibinfo{year}{2008}).

\bibitem[{\citenamefont{{Walls} and {Milburn}}(1994)}]{QO}
\bibinfo{author}{\bibfnamefont{D.~F.} \bibnamefont{{Walls}}} \bibnamefont{and}
  \bibinfo{author}{\bibfnamefont{G.~J.} \bibnamefont{{Milburn}}},
  \emph{\bibinfo{title}{Quantum Optics}} (\bibinfo{publisher}{Springer-Verlag,
  Berlin}, \bibinfo{year}{1994}).

\bibitem[{\citenamefont{Zhao et~al.}(2009)\citenamefont{Zhao, Ju, Miao, Gras,
  Fan, and Blair}}]{ZJ+09}
\bibinfo{author}{\bibfnamefont{C.}~\bibnamefont{Zhao}},
  \bibinfo{author}{\bibfnamefont{L.}~\bibnamefont{Ju}},
  \bibinfo{author}{\bibfnamefont{H.}~\bibnamefont{Miao}},
  \bibinfo{author}{\bibfnamefont{S.}~\bibnamefont{Gras}},
  \bibinfo{author}{\bibfnamefont{Y.}~\bibnamefont{Fan}}, \bibnamefont{and}
  \bibinfo{author}{\bibfnamefont{D.~G.} \bibnamefont{Blair}},
  \bibinfo{journal}{Phys. Rev. Lett.} \textbf{\bibinfo{volume}{102}},
  \bibinfo{pages}{243902} (\bibinfo{year}{2009}).

\end{thebibliography}

\end{document}